\documentstyle[12pt, fleqn]{article}
\newcommand{\ymh}{Yang-Mills-Higgs lagrangian$\;$}
\newcommand{\p}{\partial}
\newcommand{\nonum}{\nonumber\\}
\newcommand{\be}{\begin{equation}}
\newcommand{\ee}{\end{equation}}
\newcommand{\ba}{\begin{eqnarray}}
\newcommand{\ea}{\end{eqnarray}}
\newcommand{\bd}{\begin{displaymath}}
\newcommand{\ed}{\end{displaymath}}

\begin{document}
\thispagestyle{empty}
\begin{center}
\vspace*{1.0cm}
{\Large \bf Modified Reconstruction of Standard Model
\vskip 2mm
in Non-Commutative Differential Geometry}
\vskip 2.5cm
{
\Large
Yoshitaka {\sc Okumura}\footnote{
e-mail address: okum@isc.chubu.ac.jp
},
Shin-ichi {\sc Suzuki},
}
\vskip 0.2cm
{\it Department of Natural Science, Chubu University, Kasugai 487, Japan}
\vskip 2mm
and
\vskip 2mm
{
\large
Katsusada {\sc Morita}\footnote{e-mail address: h44753a@nucc.cc.nagoya-u.ac.jp}
}
\vskip 4mm
{\it Department of Physics, Nagoya University, Nagoya 464-01, Japan}
\vskip 2mm
\end{center}
\vspace{2.5 cm}
\begin{abstract}
Sogami recently proposed the new idea to express Higgs particle as a kind of
gauge particle by prescribing the generalized covariant derivative
with gauge and Higgs fields operating on quark and lepton fields.
The field strengths for both the gauge and Higgs fields are defined
by the commutators of the covariant derivative by which
he could obtain the Yang-Mills Higgs Lagrangian in the standard model.
Inspired by Sogami's work, we present a modification
of our previous scheme to formulate the spontaneously broken gauge theory
in non-commutative geometry on the discrete space $M_4\times Z_2$
  by introducing the generation mixing matrix
$K$ in $d_\chi$ operation on the fields $a_i(x,y)$ which compose the gauge and
Higgs fields. The standard model is reconstructed according to the modified
scheme, which
does not yields not only
  any special relations between the particle masses but also
  the special restriction on the Higgs potential.
\end{abstract}
\vfill\eject
\setlength{\baselineskip}{0.65cm}

\section*{\large \bf {{\mbox \S 1} Introduction}}

Many works have been done to realize the idea that the Higgs particle is
a kind of the gauge field in non-commutative geometry(NCG) on the discrete
space \cite{Con}$\sim$\cite{Naka}.
Sogami recently proposed the new idea \cite{Soga}
to express Higgs particle as a kind of
gauge particle by prescribing the generalized covariant derivative
with gauge and Higgs fields operating on quark and lepton fields.
The field strengths for both the gauge and Higgs fields are defined
by the commutators of the covariant derivative by which
he could obtain the Yang-Mills Higgs Lagrangian in the standard model
with the extra restriction on the coupling constant of the Higgs potential.
The generation mixing is from the outset considered by setting up
the interactions between fermions ( lepton, up and down quarks ) and
the Higgs field with the Yukawa couplings in the matrix form. The coupling
constant of the Higgs potential term is denoted by
three Yukawa coupling constants which yields the relation of the Higgs and top
quark masses $M_{\mathop{}_{H}}=\sqrt{2}m_t$ through
the top quark mass dominance in the trace of the product of mass matrices.
\par
The present authors  have also developed the formalism \cite{MO1}, \cite{MO2}
to be applicable to
the gauge theory with complex symmetry breaking pattern such as SU(5)
GUT \cite{MO2}, \cite{MO3} or SO(10) GUT \cite{O10}.
However, the incorporation of generation mixing was not sufficient
in our previous formalism though it was treated in the second reference
of \cite{MO1}. Inspired by th Sogami's work, we will in this paper introduce
the generation mixing matrix $K$ in the $d_\chi$ operation on $a_i(x,y)$
which composes the gauge and Higgs fields together with $M(y)$ matrix
to cause the spontaneous symmetry breakdown. $K$ was originally introduced
by Chamseddine et.al. \cite{Cham} to keep the meaningful Higgs potential.
In our formalism the Higgs potential is kept meaningful even if the generation
mixing would not exist and so we here introduce $K$ not to keep the
Higgs potential but to obtain the realistic interactions between the quark and
Higgs particle. \par
This paper is divided into four sections. The next section presents
the modifications of our previous formalism
based on the generalized differential calculus on
$M_4\times Z_2$ so as to incorporate the generation mixing mechanism and color
symmetry.
In the second reference of \cite{MO1},
$M_4\times Z_3$ was necessary to take account of
color symmetry  responsible for strong interaction. However, by considering
the generalized gauge field with direct product form we will only need
$M_4\times Z_2$ in this paper.
In this section  a geometrical
picture for the unification of the gauge and Higgs fields is realized,
which is the ultimate understanding in this field.
The third section is the application to the standard model
which leads to the quite different predictions for particle masses
from the Sogami'one.
 The last section is devoted to concluding remarks.
\section*{\large \bf {{\mbox \S 2} Generalized gauge field
                                        with direct product form}}
This section is mainly the review of our formulation
to construct the gauge theory in non-commutative geometry
on the discrete space \cite{MO1}
in which the extra discrete space $Z_3$ was necessary
to incorporate the strong interaction. We propose the modification to afford
the direct product gauge group such as SU(3)$_c\times$SU(2)$_{\mathop{}_{L}}$
which enables us to use the discrete space $M_4\times Z_2$
in reconstructing the full standard model.\par
In addition the generation mixing matrix $K$ is introduced
to accord the Sogami's idea. $K$ was initially considered in Ref.\cite{Cham}
to ensure the meaningful Higgs potential whereas our formulation did not need
$K$ to make the consistent gauge model with the spontaneous symmetry breakdown.
However, we here introduce $K$ to obtain the realistic Dirac Lagrangian with
the generation mixing.
\par
Let us first summarize the story of Ref.\cite{MO1} though we modify
it in such a way to include the generation mixing matrix $K(y)$.
The generalized gauge field $A(x,y)$ in non-commutative geometry
on the discrete space $M_4\times Z_2$ was given  as
\be
      A(x,y)=\sum_{i}a^\dagger_{i}(x,y){\bf d}a_i(x,y),\label{2.1}
\ee
where $a_i(x,y)$ is the square-matrix-valued function
and ${\bf d}$ is the generalized exterior derivative defined as follows.
\ba
&&       {\bf d}a_{i}(x,y)=(d + d_\chi)a(x,y)
                                   =(d+d_{\chi})a(x,y), \nonum
&&     da_i(x,y) = \partial_\mu a_i(x,y)dx^\mu,\hskip 1cm \nonum
&&   d_{\chi} a_i(x,y) =K(y)[-a_i(x,y)M(y) + M(y)a_i(x,-y)]\chi.
        \label{2.2}
\ea
Here
$dx^\mu$ is
ordinary one form basis, taken to be dimensionless, in $M_4$, and $\chi$
is the one form basis, assumed to be also dimensionless,
in the discrete space $Z_2$.
We have introduced $x$-independent matrix $M(y)$
whose hermitian conjugation is given by $M(y)^\dagger=M(-y)$.
$K(y)$ is also assumed to be $K(y)^\dagger=K(-y)$
and commutes with $a_i(x,y)$ and $M(y)$.
We here skip to explain detailed algebras
with respect to non-commutative geometry
because those are seen in Ref.\cite{MO1}.
According to Ref.\cite{MO1}, we can define
the gauge fields $A_\mu(x,y)$ and the Higgs field $\Phi(x,y)$ as
\ba
&&    A_\mu(x,y) = \sum_{i}a_{i}^\dagger(x,y)\p_\mu a_{i}(x,y),  \nonum
&&\Phi(x,y) = \sum_{i}
a_{i}^\dagger(x,y)\,(-a_i(x,y)M(y)
            + M(y)a_i(x,-y)),
  \label{2.3}
\ea
with which Eq.(\ref{2.1}) is rewritten as
\be
      A(x,y)=A_\mu(x,y)dx^\mu+K(y)\Phi(x,y)\chi. \label{2.E1}
\ee
In connection with $K(y)$
in Eq.(\ref{2.2}), it should be noticed
that $a_i(x,y)$ is also a representation
in the generation space and so does $A(x,y)$ in Eq.(\ref{2.E1}).
Eq.(\ref{2.E1}) expresses the unified picture of the gauge and Higgs fields as
the generalized connection on the discrete space $M_4\times Z_2$.
\par
We extend Eq.(\ref{2.1}) to the generalized gauge field
with the direct product form to
incorporate gluon field on the same sheet as flavor gauge fields and to contain
the generation mixing matrix $K(y)$ to accord the Sogami's idea.
\be
      {\cal A}(x,y)=\sum_{i}a^\dagger_{i}(x,y){\bf d}a_i(x,y)\otimes {\bf 1}
      +{\bf 1} \otimes \sum_{j}b^\dagger_{j}(x,y){\bf d}b_j(x,y),
      \label{2.4}
\ee
where the second term is responsible for the gluon field,
 so that actually  ${\bf d}b_j(x,y)={d}b_j(x,y)$
 because the strong interaction does not break down spontaneously,
 and we denote $b_j(x,y)=b_j(x)$  which means the strong interaction
 works on both discrete spaces $(y=\pm)$.
In the same context as in Eq.(\ref{2.3}), the gluon field $G^{}_\mu(x)$
is expressed as
\be
   G^{}_\mu(x)=\sum_{j}b_{j}^\dagger(x)\partial_\mu b_{j}(x).
   \label{2.5}
\ee
In order to identify  $A_\mu(x,y)$ and $G^{}_\mu(x)$ as true gauge fields,
the following conditions have to be imposed.
\ba
&&    \sum_{i}a_{i}^\dagger(x,y)a_{i}(x,y)= 1,  \nonum
&&     \sum_{j}b_{j}^\dagger(x)b_{j}(x)={1\over g_3},
 \label{2.6}
\ea
where $g_3$ is a constant related to
the corresponding coupling constant as shown later. $i$ and $j$
are  variables of the extra
internal space which we can not now identify what they are.
Eqs.(\ref{2.3}) and
(\ref{2.5}) are very similar to the effective gauge field
in Berry phase \cite{Ber}, which may
lead to the identification of this internal space.
In general, we can put the right hand side of the first equation
in Eq.(\ref{2.6})
to be ${1/g_y}$. However, we put it as it is to avoid the complexity.
\par
Before constructing the gauge covariant field strength,
we address the gauge transformation
of $a_i(x,y)$ and $b_j(x)$ which is defined as
\ba
&&      a^{g}_{i}(x,y)= a_{i}(x,y)g(x,y), \nonum
&&      b^{g}_{j}(x) =  b_j(x)g_3(x),
\label{2.7}
\ea
where
$g(x,y)$ and $g_3(x)$ are the gauge functions with respect to the corresponding
flavor unitary group and the color SU(3)$_c$ group, respectively.
Then, we can get the gauge transformation of ${\cal A}(x,y)$ to be
\ba
{\cal A}^g(x,y)&=&g^{-1}(x,y)\otimes g^{-1}_3(x){\cal A}(x,y)g(x,y)\otimes
g_3(x)\nonum
&&+g^{-1}(x,y){\bf d}g(x,y)\otimes {\bf 1}+{\bf 1}\otimes {1\over g_3}
\,g^{-1}_3(x){d}g_3(x),
 \label{2.8}
\ea
where use has been made of Eq.(\ref{2.6}) and  as in Eq.(\ref{2.2}),
\be
       {\bf d}g(x,y)=
      \partial_\mu g(x,y)dx^\mu
                + K(y)[-g(x,y)M(y) + M(y)g(x,y)]\chi.
        \label{2.9}
\ee
Eq.(\ref{2.8}) affords us
to construct the gauge covariant field strength as follows:
\be
  {\cal F}(x,y)= F(x,y)\otimes {\bf 1} + {\bf 1} \otimes {\cal G}(x),
\label{2.10}
\ee
where $F(x,y)$ and ${\cal G}(x)$ are the field strengths
of flavor and color gauge fields, respectively and given as
\ba
&&     F(x,y) = {\bf d}A(x,y)+A(x,y)\wedge A(x,y),     \nonum
&&     {\cal G}(x)   =d\,G^{}(x)+g_3 G^{}(x)\wedge G^{}(x).
\label{2.11}
\ea
The algebras of non-commutative differential geometry defined in Ref.\cite{MO1}
 yields
\ba
 F(x,y) &=& { 1 \over 2}F_{\mu\nu}(x,y)dx^\mu \wedge dx^\nu  \nonum
           &&\hskip 1.5cm   + K(y)D_\mu \Phi(x,y)dx^\mu \wedge \chi
               + K(y)K(-y)V(x,y)\chi \wedge \chi,
                \label{2.12}
\ea
where
\ba
&&  F_{\mu\nu}(x,y)=\p_\mu A_\nu (x,y) - \p_\nu A_\mu (x,y)
               +[A_\mu(x,y), A_\mu(x,y)],\nonum
&&  D_\mu \Phi(x,y)=\p_\mu \Phi(x,y)  + A_\mu(x,y)(M(y) + \Phi(x,y))\nonum
&&     \hskip 6.5cm            -(\Phi(x,y)+M(y))A_\mu(x,-y),\nonum
&&  V(x,y)= (\Phi(x,y) + M(y))(\Phi(x,-y) + M(-y)) - Y(x,y). \label{2.13}
\ea
$Y(x,y)$ in Eq.(\ref{2.13}) is auxiliary field and expressed as
\be
  Y(x,y)= \sum_{i}a_{i}^\dagger(x,y)M(y)M(-y)a_{i}(x,y),
 \label{2.14}
\ee
which may be independent or dependent of $\Phi(x,y)$
and/or may be a constant field.
If we define $H(x,y) = \Phi(x,y) + M(y),$
it is readily known that
the function $H(x,y)$ represents the unshifted Higgs field,
whereas $\Phi(x,y)$ denotes the shifted Higgs field
with vanishing vacuum expectation value so
that $M(y)$ determines the scale and pattern of the
spontaneous breakdown of gauge symmetry.
In contrast to $F(x,y)$, ${\cal G}(x)$ is simply denoted as
\ba
 {\cal G}(x)&=&{1\over 2}{G}_{\mu\nu}(x)dx^\mu\wedge dx^\nu \nonum
        &=&{1\over 2}\{\partial_\mu G^{}_\nu(x)-\partial_\nu G^{}_\mu(x)
        + g_3[G^{}_\mu(x), G^{}_\mu(x)]\}dx^\mu\wedge dx^\nu.
\label{2.15}
\ea
\par
With the same metric structure on the discrete space
$M_4\times Z_{\mathop{}_{N}}$ as in Ref.\cite{MO1}
 we can obtain the gauge invariant
\ymh(YMH)
\ba
{\cal L}_{{\mathop{}_{YMH}}}(x)&=&-{\rm Tr}\sum_{y=\pm}{1 \over g_{y}^2}
< {\cal F}(x,y),  {\cal F}(x,y)>\nonum
&=&-{\rm Tr}\sum_{y=\pm}{1\over 2g^2_y}
F_{\mu\nu}^{\dag}(x,y)F^{\mu\nu}(x,y)\nonum
&&+{\rm Tr}\sum_{y=\pm}{\alpha^2\over g_{y}^2}[K(-y)K(y)]
    (D_\mu \Phi(x,y))^{\dag}D^\mu \Phi(x,y)  \nonum
&& -{\rm Tr}\sum_{y=\pm}{\beta^4\over g_{y}^2}[K(-y)K(y)]^2
        V^{\dag}(x,y)V(x,y)  \nonum
&&-{\rm Tr}\sum_{y=\pm}{1\over 2g_{y}^2}{ G}_{\mu\nu}^{\dag}(x)
{ G}^{\mu\nu}(x),
\label{2.16}
\ea
where $g_y$ is a constant relating
to the coupling constant of the flavor gauge field and
Tr denotes the trace over internal symmetry matrices including the color,
flavor symmetries and generation space.
$\alpha$ and $\beta$ emerge from
the definition of metric $<\chi, \chi>=-\alpha^2 $ and
$<\chi\wedge\chi, \chi\wedge\chi>=\beta^4 $, respectively.
The third term in the right hand side is the potential term of Higgs particle.
\par
Let us turn to the fermion sector to construct the Dirac Lagrangian. This is
also deeply indebted to Ref.\cite{MO1} so that only main points
should be explained by skipping details.
Let us start to define the covariant derivative acting
on the spinor field $\psi(x,y)$ which is the representation
of the corresponding semi simple group including SU(3)$_c$.
\be
{\cal D}\psi(x,y)=({\bf d}+ {\cal A}(x,y))\psi(x,y),
\label{2.17}
\ee
which we call the covariant spinor one-form.
The algebraic rules in Ref.\cite{MO1} along with Eq.(\ref{2.4})
leads Eq.(\ref{2.17}) to
\ba
{\cal D}\psi(x,y)&=& \{{\bf 1}\otimes {\bf 1}\p_\mu dx^\mu
+ (A^f_\mu(x,y)\otimes {\bf 1}dx^\mu
      + K(y)H(x,y)\otimes {\bf 1}\chi) \nonum
 &&+{\bf 1}\otimes G^{f}_\mu(x)dx^\mu\}\psi(x,y),
\label{2.18}
\ea
where $A^f_\mu(x,y)$ and $G^{f}_\mu(x)$ are the differential representations
with respect to $\psi(x,y)$.
It should be noticed
that ${\cal D}\psi(x,y)$ is gauge covariant so that
\be
     {\cal D}\psi^g(x,y)=(g^f(x,y))^{-1}\otimes
     {(g^f_s(x))}^{-1}{\cal D}\psi(x,y), \label{2.19}
\ee
where
$g^f(x,y)\otimes g^f_s(x)$ is the gauge transformation function with respect to
the representation of $\psi(x,y)$.
Corresponding with Eq.(\ref{2.17}),
the associated spinor one-form is introduced by
\be
{\tilde {\cal D}}\psi(x,y)= {\bf 1}\otimes {\bf 1}\{\gamma_\mu \psi(x,y)dx^\mu
              -i{c_{ \mathop{}_{Y}}}\psi(x,y)\chi\},
\label{2.20}
\ee
where $c_{ \mathop{}_{Y}}$ is a real dimensionless constant related to the
Yukawa coupling constant between Higgs field and fermions. With the same inner
products for spinor one-forms as in Ref.\cite{MO1},
 we can get the Dirac Lagrangian.
\ba
{\cal L}_{ \mathop{}_{D}}(x,y)
     &=&i{\rm Tr}<{\tilde {\cal D}}\psi(x,y),{\cal D}\psi(x,y)>\nonum
&=&i\,{\rm Tr}\,[\,{\bar\psi}(x,y)\gamma^\mu({\bf 1}\otimes{\bf 1}\partial_\mu
+A_\mu^{f}(x,y)\otimes {\bf 1}+ {\bf 1}\otimes G^{f}_\mu(x))\psi(x,y) \nonum
&&+i{c_{ \mathop{}_{Y}}}\alpha^2{\bar\psi}(x,y)\sum_{y=\pm}
                                    K(y)H(x,y)\otimes{\bf 1}\psi(x,y)\,],
\label{2.21}
\ea
where
Tr is also the trace over internal symmetry matrices including the color,
flavor symmetries and generation space.
The total Dirac Lagrangian is the sum over $y$:
\be
{\cal L}_{\mathop{}_{D}}(x)
      =\sum_{y=\pm}{\cal L}_{\mathop{}_{D}}(x,y),
\label{2.22}
\ee
which is apparently invariant for the Lorentz and gauge transformations.
Eqs.(\ref{2.16}) and (\ref{2.22}) along with Eq.(\ref{2.21})
 are crucially important to reconstruct the
spontaneously broken gauge theory.
\par
With these preparations, we can apply the direct product formalism proposed
in this section to the standard model
and compare it with the Sogami's presentation \cite{Soga}.

\section*{\large \bf {{\mbox \S 3} Model Construction}}

We first prescribe the fermion field $\psi(x,y)$ in Eq.(\ref{2.17})
with the existing leptons and quarks
and then decide the generalized gauge field ${\cal A}(x,y)$
in order to give the correct Dirac Lagrangian for the fermion sector
in the standard model. Hereafter, the argument $x$ is often abbreviated
if no confusion.
\be
       \psi(x,+)=\left(\matrix{ l_{ \mathop{}_{L}}\cr
                            \gamma q_{ \mathop{}_{L}}\cr
                            \sqrt{1-\gamma^2} q_{ \mathop{}_{L}}\cr}
                            \right),
\hskip 1.5cm
       \psi(x,-)=\left(\matrix{ e_{ \mathop{}_{R}}\cr
                               d_{ \mathop{}_{R}}\cr
                                  u_{ \mathop{}_{R}}\cr}
                            \right), \label{3.24}
\ee
where $l_{ \mathop{}_{L}}$ and $q_{ \mathop{}_{L}}$ are the left-handed
doublet lepton and quark, respectively and $\gamma$ is a constant
necessary for the
normalization of the kinetic term of $q_{ \mathop{}_{L}}$.
It should be noticed that $\psi(x,y)$ has the index for the three generation
and so do the explicit expressions for fermions in the right hand sides
of Eq.(\ref{3.24}). For example, in the strict expressions
 $e_{ \mathop{}_{R}},$ $d_{ \mathop{}_{R}},$ and $u_{ \mathop{}_{R}}$
 in the right of Eq.(\ref{3.24}) should be written as
\be
       e_{ \mathop{}_{R}}\to \left(\matrix{ e_{ \mathop{}_{R}}\cr
                               \mu_{ \mathop{}_{R}}\cr
                                  \tau_{ \mathop{}_{R}}\cr}
                            \right), \hskip 1cm
       d_{ \mathop{}_{R}}\to \left(\matrix{ d_{ \mathop{}_{R}}\cr
                               s_{ \mathop{}_{R}}\cr
                                  b_{ \mathop{}_{R}}\cr}
                            \right), \hskip 1cm
       u_{ \mathop{}_{R}}\to \left(\matrix{ u_{ \mathop{}_{R}}\cr
                               c_{ \mathop{}_{R}}\cr
                                  t_{ \mathop{}_{R}}\cr}
                            \right),
                            \label{3.s1}
\ee
respectively.
\par
In order to obtain the Dirac Lagrangian for fermion fields in Eq.(\ref{3.24})
we denote  the generalized gauge field ${\cal A}(x,y)$
in Eq.(\ref{2.4}) as follows:
\be
      {\cal A}(x,y)={\cal A}_\mu(x,y)dx^\mu\otimes {\bf 1}
               + {\Phi}^{'}(x,y)\chi\otimes {\bf 1}
               + {\bf 1}\otimes {\cal G}_\mu(x)dx^\mu. \label{3.25}
\ee
${\cal A}_\mu(x,\pm)$  are specified as
\ba
    && {\cal A}_\mu(x,+)=-\frac i2\sum_{k=1}^3\tau^k A_\mu^k
                          -\frac i2a\tau^0B_\mu, \label{3.26} \\
    && {\cal A}_\mu(x,-)=-\frac i2bB_\mu,         \label{3.27}
\ea
where
$A_\mu^i$ and $B_\mu$ are SU(2) and U(1) gauge fields, respectively
and so $\tau^i$ is the Pauli matrices and $\tau^0$ is $2\times 2$ unit matrix.
$a$ and $b$ in Eqs.(\ref{3.26}) and (\ref{3.27})
are the U(1) hypercharge matrices
corresponding to Eq.(\ref{3.24}) and expressed as
\be
     a=\left(\matrix{ -1 & 0 & 0 \cr
                      0 & \frac13 & \cr
                      0 & 0 & \frac13\cr}
                      \right),
                      \hskip 2cm
     b=\left(\matrix{ -2 & 0 & 0 \cr
                      0 & -\frac23 & \cr
                      0 & 0 & \frac43\cr}
                      \right).     \label{3.28}
\ee
${\Phi}^{'}(x,y)$  is also written  in accord with
Eq.(\ref{3.24}) as
\be
            {\Phi}^{'}(x,y)
                  =\left(\matrix{ \Phi(x,y) & 0 & 0 \cr
                              0  & \Phi(x,y)  & 0 \cr
    0  &   0   &{\tilde \Phi}(x,y)  \cr}\right),\label{3.29}
\ee
where
\ba
  &&\Phi(x,+)=\Phi(x,-)^\dagger=\left(\matrix{ \phi^+ \cr
                           \phi^0 \cr}\right),
                   \hskip 1cm        {\tilde \Phi}(x,+)=
                   {\tilde \Phi}(x,-)^\dagger=i\tau^2\Phi^\ast(x,+),
                           \label{3.30} \\
       &&       M(+)=M(-)^\dagger
                                             =\left(\matrix{ 0 \cr
                                               \mu  \cr}\right),
                   \hskip 1cm {\tilde M}(+)={\tilde M}(-)^\dagger
                                 =i\tau^2 {M}(+)
                                             =\left(\matrix{ \mu \cr
                                                   0  \cr}\right).
                                               \label{3.31}
\ea
 and  ${\cal G}_\mu(x)$ is expressed as
\be
            {\cal G}_\mu(x)
                    =\left(\matrix{ 0 & 0 & 0 \cr
                                   0  & G_\mu(x) & 0 \cr
                             0  &   0   & G_\mu(x)\cr}\right), \label{3.32}
\ee
with $G_\mu(x)$  written as
\be
         G_\mu(x)=-\frac i2\sum_{a=1}^8\lambda^a G_\mu^a, \label{3.33}
\ee
where $G_\mu^a$ is SU(3) color gauge field
and so $\lambda^a$ are Gell-Mann matrices.\par
With these specifications, the generalized gauge fields ${\cal F}(x,y)$
\ba
{\cal F}(x,y)&=&\frac12{\cal F}_{\mu\nu}(x,y)\otimes{\bf 1}dx^\mu\wedge dx^\nu
   +{\cal K}(y)D_\mu {\cal H}(x,y)\otimes{\bf 1}dx^\mu\wedge\chi \nonum
   &&+{\cal K}(y){\cal K}(-y){\cal V}(x,y)\otimes{\bf 1}\chi\wedge\chi
   +{\bf 1}\otimes \frac12{\cal G}_{\mu\nu}(x)dx^\mu\wedge dx^\nu
   \label{3.34}
\ea
can be determined as
\ba
&&  \!\!\!{\cal F}_{\mu\nu}(x,+)=-\frac i2\sum_{i=1}^3\tau^i\left(
        \p_\mu A_\nu^i-\p_\mu A_\nu^i+\epsilon_{ijk}A_\mu^jA_\nu^k\right)
        -\frac i2\tau^0 a\left(\p_\mu B_\nu-\p_\nu B_\mu \right),\label{3.35}\\
&&  \!\!\!{\cal F}_{\mu\nu}(x,-)=-\frac i2\tau^0 b
                 \left(\p_\mu B_\nu-\p_\nu B_\mu \right),  \label{3.36} \\
&&  \!\!\! D_\mu {\cal H}(x,+)=\left(D_\mu {\cal H}(x,-)\right)^\dagger
= \left(\matrix{D_\mu H & 0 & 0 \cr
                        0 & D_\mu H &  0\cr
                        0 & 0 & D_\mu {\tilde H} \cr} \right) \label{3.37}
\ea
with
\ba
  &&    D_\mu H=\p_\mu\Phi - \frac i2\sum_{i=1}^3\tau^iA_\mu^i(\Phi+M)
                          -\frac i2\tau^0B_\mu(\Phi+M), \label{3.38}\\
  &&    D_\mu {\tilde H}=\p_\mu{\tilde\Phi} - \frac i2\sum_{i=1}^3\tau^iA_\mu^i
            ({\tilde\Phi}+{\tilde M})
                  +\frac i2\tau^0B_\mu({\tilde\Phi}+{\tilde M}), \label{3.39}
\ea
and
\ba
 &&  {\cal V}(x,y)=\left(\matrix{V(x,y) & 0 & 0 \cr
                                  0  & V(x,y)& 0 \cr
                                  0 & 0 & {\tilde V}(x,y)\cr}\right),
                             \label{3.40}
\ea
with
\ba
 \hskip -1cm V(x,+)=(\Phi+M)(\Phi^\dagger+M^\dagger)-Y(+),&&
 V(x,-)=(\Phi+M)^\dagger(\Phi+M)-Y(-),
    \label{3.41}\\
 \hskip -1cm {\tilde V}(x,+)=({\tilde\Phi}+{\tilde M})({\tilde\Phi}^\dagger
+{\tilde M}^\dagger)-{\tilde Y}(+), &&
 {\tilde V}(x,-)=({\tilde\Phi}+{\tilde M})^\dagger({\tilde\Phi}
+{\tilde M})-{\tilde Y}(-),
\label{3.42}
\ea
and in addition to these expressions the generation mixing matrix
${\cal K}(y)$ has the following form corresponding to Eq.(\ref{3.24}).
\ba
&&{\cal K}(+)={\cal K}(-)^\dagger={\cal K}=
                    \left(\matrix{ K^l & 0 & 0 \cr
                        0 & K^d  &  0\cr
                        0 & 0 & K^u \cr} \right), \label{3.43}
\ea
where $K^l,$ $K^d$ and $K^u$ are mixing matrices
for the lepton, down and up quarks, respectively. ${\cal G}_{\mu\nu}(x)$
in Eq.(\ref{3.34}) is given as
\ba
 &&  {\cal G}_{\mu\nu}(x)=\left(\matrix{0 & 0 & 0 \cr
                                  0  & G_{\mu\nu}& 0 \cr
                                  0 & 0 & G_{\mu\nu}\cr}\right),
                             \label{3.44}
\ea
with  $G_{\mu\nu}$ in Eq.(\ref{2.15}).
\par
Putting Eq.(\ref{3.34}) into Eq.(\ref{2.16})
and rescaling  gauge and Higgs fields
we can obtain YMH for the standard model
as follows:
\ba
{\cal L}_{{\mathop{}_{YMH}}}&=&
   -\frac14\sum_{i=1}^3\left(F_{\mu\nu}^i\right)^2
   -\frac14B_{\mu\nu}^2 \nonum
  &&  +|D_\mu H|^2   -\lambda(H^\dagger H-{\mu'}^2)^2 \nonum
 && - \frac{1}{4}
      \sum_{a=1}^8{G^a_{\mu\nu}}^{\dagger}{G^a}^{\mu\nu}, \label{3.45}
\ea
where
\ba
   &&   F_{\mu\nu}^i=\partial_\mu A_\nu^i-\partial_\nu A_\mu^i
         +g\epsilon^{ijk}A_\mu^jA_\nu^k,  \nonum
   &&   B_{\mu\nu}=\partial_\mu B_\nu-\partial_\nu B_\mu,\nonum
   &&     D^\mu H=[\,\p_\mu-{i\over 2}\,(\sum_i\tau^igA_\mu^i
          + \,\tau^0\,g'B_\mu\,)\,]\,(\Phi+M), \nonum
   &&   G_{\mu\nu}^a=\partial_\mu G_\nu^a-\partial_\nu G_\mu^a
         +g_sf^{abc}G_\mu^bG_\nu^c, \label{3.46}
\ea
with
\ba
 &&     g^2=\frac{g_+^2}{21}, \hskip 1cm
      {g'}^2=\frac{g_+^2g_-^2}{16g_+^2+5g_-^2}, \label{3.47}\\
 &&     \lambda=\frac{g_+^4g_-^2\beta^4{\rm Tr}({\cal K}{\cal K}^\dagger)^2}
         {\alpha^4(g_+^2+g_-^2)^2({\rm Tr}({\cal K}{\cal K}^\dagger))^2},
         \hskip 1cm
    {\mu'}^2= (\frac1{g_+^2}+\frac1{g_-^2})
    {\rm Tr}({\cal K}{\cal K}^\dagger)\alpha^2\mu^2,
         \label{3.48}\\
 &&        g_s^2=\frac{g_+^2g_-^2g_3^2}{2(g_+^2+g_-^2)}.  \label{3.49}
\ea
${\rm Tr}({\cal K}{\cal K}^\dagger)$ and
${\rm Tr}({\cal K}{\cal K}^\dagger)^2$ in above equations are given as
\ba
&&     {\rm Tr}({\cal K}{\cal K}^\dagger)=
    {\rm tr}( K^l{K^l}^\dagger)+3{\rm tr}( K^d{K^d}^\dagger)
               +3{\rm tr}( K^u{K^u}^\dagger),\nonum
&&     {\rm Tr}({\cal K}{\cal K}^\dagger)^2=
         {\rm tr}({K^l}{K^l}^\dagger)^2+3{\rm tr}({K^d}{K^d}^\dagger)^2
            +3{\rm tr}({K^u}{K^u}^\dagger)^2, \label{3.50}
\ea
where tr is the trace over the generation space and the factor 3  comes from
the trace of color indices.
In deriving Eq.(\ref{3.45}) the Higgs potential term $V(x,+)$ is eliminated
because the auxiliary field $Y(x,+)$ is independent field
owing to Eq.(\ref{3.31}) whereas $V(x,-)$ remains thanks to $Y(x,-)=\mu^2$.
Eq.(\ref{3.47}) yields the Weinberg angle with the parameter
$\delta={g_+}/{g_-}$ to be
\be
          \sin^2\theta_{ \mathop{}_{W}}=\frac{21}{16\delta^2+26}, \label{3.51}
\ee
and Eq.(\ref{3.48}) results in
\ba
 &&       m_{ \mathop{}_{W}}=\sqrt{\frac{1+\delta^2}2}
                 \left(\frac{{\rm Tr}({\cal K}{\cal K}^\dagger)}{21}
                           \right)^{\frac12}\alpha\mu,
                                                              \label{3.52}\\
 &&       m_{ \mathop{}_{H}}= \frac{{2}\delta\epsilon}{\sqrt{1+\delta^2}}
                    \left(\frac{{\rm Tr}({\cal K}{\cal K}^\dagger)^2}
                    {{\rm Tr}({\cal K}{\cal K}^\dagger)}
                    \right)^{\frac12}\alpha\mu,                   \label{3.53}
\ea
where $\epsilon=\beta^2/\alpha^2$. \par
These estimations are only valid in the classical level. Though we are tempted
to compare them with the experimental values
we will learn  it to be impossible after getting the Dirac lagrangian
in the fermion sector.\par
Let us turn to the construction of the Dirac Lagrangian for fermion sector.
After the rescaling of the boson fields, we can write the covariant spinor
one-form in Eq.(\ref{2.18})
corresponding with the specification of Eq.(\ref{3.24}) as
\ba
{\cal D}\psi(x,+)&=& {\bf 1}\otimes {\bf 1}\p_\mu dx^\mu
 +\{-\frac i2 (g\sum_{i=1}^3\tau^iA_\mu^i+ag'\tau^0B_\mu)\psi(x,+)dx^\mu
  \nonum
   && + {\cal K}(\Phi'+M')\psi(x,-)\chi\}\otimes {\bf 1}
 -{\bf 1}\otimes\frac i2
    \sum_{a=1}^8\lambda^ag_c{\cal G}^a_\mu\psi(x,+) dx^\mu,
              \label{3.54}
\ea
and
\ba
{\cal D}\psi(x,-)&=& {\bf 1}\otimes {\bf 1}\p_\mu dx^\mu
 +\{-\frac i2 bg'B_\mu\psi(x,-)dx^\mu
    + {\cal K}^\dagger(\Phi'+M')^\dagger\psi(x,+)\chi\}\otimes {\bf 1} \nonum
   && -{\bf 1}\otimes \frac i2\sum_{a=1}^8\lambda^ag_c{\cal G}^a_\mu
   \psi(x,-) dx^\mu. \label{3.55}
\ea
We can also express the associated spinor one-form in Eq.(\ref{2.20}) as
\be
{\tilde {\cal D}}\psi(x,\pm)= {\bf 1}\otimes {\bf 1}
\{\gamma_\mu \psi(x,\pm)dx^\mu
              -i{c_{ \mathop{}_{Y}}}\psi(x,\pm)\chi\},
\label{3.56}
\ee
where
corresponding to the expression Eq.(\ref{3.24}) $c_{ \mathop{}_{Y}}$
has the following form:
\be
           c_{ \mathop{}_{Y}}=
                    \left(\matrix{ c^l & 0 & 0 \cr
                        0 & c^d  &  0\cr
                        0 & 0 & c^u \cr} \right)
                        . \label{3.57}
\ee
$c^l$, $c^d$ and $c^u$ in Eq.(\ref{3.57}) may be matrices
in the generation space. According to Eqs.(\ref{2.21}) and (\ref{2.22}), we can
get the Dirac lagrangian
for the standard model as follows:
\ba
{\cal L}_{ \mathop{}_{D}}&=&\sum_{y=\pm}
     i<{\tilde {\cal D}}\psi(x,y),{\cal D}\psi(x,y)>\nonum
  &=& i
  \left( {\bar l}_{ \mathop{}_{L}},\,
                            \gamma {\bar q}_{ \mathop{}_{L}},\,
                            \sqrt{1-\gamma^2} {\bar q}_{ \mathop{}_{L}}
                            \right)
                            \left\{\p_\mu-\frac i2\sum_{a=1}^8
                            \lambda^ag_c{\cal G}^a_\mu-\frac i2\sum_{i=1}^3
                            \tau^ig{\cal A}^i_\mu\right.\nonum
                 &&  \hskip 5cm   \left.-\frac i2\tau^0
                    \left(\matrix{ -1 & 0 & 0 \cr
                        0 & \frac13  &  0\cr
                        0 & 0 & \frac13 \cr} \right)g'B_\mu \right\}
       \left(\matrix{
       l_{ \mathop{}_{L}}\cr
                            \gamma q_{ \mathop{}_{L}}\cr
                            \sqrt{1-\gamma^2} q_{ \mathop{}_{L}}\cr}
                            \right)\nonum
  &&+i
  \left({\bar e}_{\mathop{}_{R}},\,
                             {\bar d}_{ \mathop{}_{R}},\,
                             {\bar u}_{ \mathop{}_{R}}
                            \right)
                            \left\{\p_\mu-\frac i2\sum_{a=1}^8
                            \lambda^ag_c{\cal G}^a_\mu
                   -\frac i2
                    \left(\matrix{ -2 & 0 & 0 \cr
                        0 & -\frac23  &  0\cr
                        0 & 0 & \frac43 \cr} \right)g'B_\mu\right\}
       \left(\matrix{
       e_{ \mathop{}_{R}}\cr
                            d_{ \mathop{}_{R}}\cr
                            u_{ \mathop{}_{R}}\cr}
                            \right)\nonum
    && -
  \left( {\bar l}_{ \mathop{}_{L}},\,
                            \gamma {\bar q}_{ \mathop{}_{L}},\,
                            \sqrt{1-\gamma^2} {\bar q}_{ \mathop{}_{L}}
                            \right)
                  g_{ \mathop{}_{Y}}\left(
                  \Phi'+M'\right)
       \left(\matrix{
       e_{ \mathop{}_{R}}\cr
                            d_{ \mathop{}_{R}}\cr
                            u_{ \mathop{}_{R}}\cr}
                            \right) \nonum
&&  -\left({\bar e}_{\mathop{}_{R}},\,
                             {\bar d}_{ \mathop{}_{R}},\,
                             {\bar u}_{ \mathop{}_{R}}
                            \right)
                  g_{ \mathop{}_{Y}}^\dagger\left(
                  \Phi'+M'\right)^\dagger
       \left(\matrix{
       l_{ \mathop{}_{L}}\cr
                            \gamma q_{ \mathop{}_{L}}\cr
                            \sqrt{1-\gamma^2} q_{ \mathop{}_{L}}
                            \cr} \right)
\label{3.58}
\ea
which is sufficient as the Dirac Lagrangian of the standard model
with the Yukawa coupling constants $g_{ \mathop{}_{Y}}$ in matrix form
given as
\be
    g_{ \mathop{}_{Y}}=c_{ \mathop{}_{Y}}\alpha^2${\cal K}$
          =\left(\matrix{ g_{ \mathop{}_{Y}}^l & 0 &  0 \cr
                       0 & g_{ \mathop{}_{Y}}^d & 0 \cr
                       0 & 0 & g_{ \mathop{}_{Y}}^u
                              \cr} \right)
          =\left(\matrix{  c^l\alpha^2K^l  & 0 & 0 \cr
                       0 & \gamma c^d\alpha^2K^d & 0 \cr
                       0 & 0 & \sqrt{1-\gamma^2}c^u\alpha^2K^u
                              \cr} \right), \label{3.59}
\ee
and yields
the fermion mass  as follows:
\ba
L_{\mbox{\tiny fermion mass}}
&=&-{\bar e}_{ \mathop{}_{L}}M^le_{ \mathop{}_{R}}-
               {\bar d}_{ \mathop{}_{L}}M^dd_{ \mathop{}_{R}}
        - {\bar u}_{ \mathop{}_{L}}M^u
                      u_{ \mathop{}_{R}} - H.C.,\label{3.60}
\ea
where
$M^l=g_{ \mathop{}_{Y}}^l\mu',$
$M^d=\gamma g_{ \mathop{}_{Y}}^d\mu',$ and
$M^u=\sqrt{1-\gamma^2}g_{ \mathop{}_{Y}}^u\mu'$
 are the mass matrices appeared in Ref.$\,$\cite{Soga}.
In our case it seems to be impossible to connect the Higgs mass
$m_{ \mathop{}_{H}}$ and the top quark mass $m_t$ because of
the fact that the top quark contribution to
${\rm Tr}({\cal K}{\cal K}^\dagger)^2$ and ${\rm Tr}({\cal K}{\cal K}^\dagger)$
in Eq.(\ref{3.53}) may not be necessarily dominant
due to its trace form itself
and so many unknown constants in Eq.(\ref{3.59}) including $c^l$, $c^d$ and
$c^u$ which may be matrices in the generation space.
Thus, we can say nothing about the relation of $m_t$ and $m_{ \mathop{}_{H}}$
written in Eq.(\ref{3.53}).
In addition, the parameter $\epsilon=\beta^2/\alpha^2$ which is amount to
the existence of the independent quadratic Higgs potential term
of $g_+$ and $g_-$ appears in
Eq(\ref{3.45}). It makes the prediction of the Higgs mass completely
ambiguous, and so it would be wise way to cease to make predictions
about the relations between these particle masses though tempting.

\section*{\large \bf {{\mbox \S 4} Conclusion}}
Inspired by the Sogami's work \cite{Soga}
we introduced the generation mixing matrix
${\cal K}$ in the $d_\chi$ operation on $a_i(x,y)$ which composes the gauge
and the Higgs particles. The estimations of
${\rm Tr}({\cal K}{\cal K}^\dagger)^2$ and ${\rm Tr}({\cal K}{\cal K}^\dagger)$
in Eq(\ref{3.52}) and Eq(\ref{3.53}) are so difficult that we can not say
anything about $m_{ \mathop{}_{H}}$ and $m_{ \mathop{}_{W}}$.
${\cal K}$ was originally introduced in \cite{Cham} to prevent the Higgs
potential terms from vanishing because of the auxiliary fields $Y_{nm}$
and keep the meaningful Higgs potential terms, however it loses meaning
in the case of one generation or no mixing between generations. Contrary to
this we can get the previous results
for $m_{ \mathop{}_{H}}$ and $m_{ \mathop{}_{W}}$
in \cite{MO1} if ${\cal K}$ is unit
matrix in the corresponding spaces and so
${\rm Tr}({\cal K}{\cal K}^\dagger)^2={\rm Tr}({\cal K}{\cal K}^\dagger)=21$.
The expression of $ \sin^2\theta_{ \mathop{}_{W}}$ in Eq.(\ref{3.51})
considerably changes due to the implicit effects from the color and
generation spaces, however it is not crucial difference to be exclusive
from each other.\par
Contrary to ours, there is a simple relation
$m_{ \mathop{}_{H}}=\sqrt{2}m_t$ in Ref.\cite{Soga}.
This is because $\lambda$ and $m_{ \mathop{}_{H}}$ are in his paper expressed
directly by the traces of Yukawa coupling constants
not by $K$ matrix appeared in our case.
There seems considerable differences
between our and Sogami's formalisms though both are based on
 the similar ideas to treat gauge and the Higgs fields on the same footing.
However, it is impossible to decide which case is better
in the present time. There are many other attractive approaches in this
field of non-commutative geometry.
It is now expected that the Higgs search will be successful in the near future.
\par
Talking about the quantization, we have the same number of parameters of
the standard model in this paper as the ordinary one.  Not only the special
relations between parameters such as coupling constants but also
the Higgs potentials with the restrictive forms are never introduced here.
Thus, the quantization can be performed in the same way as in the ordinary one.

\vskip 0.5 cm
\begin{center}
{\bf Acknowledgement}
\end{center}
The authors would like to
express their sincere thanks to
Professors J.~IIzuka,
 H.~Kase and M.~Tanaka
for useful suggestion and
invaluable discussions on the non-commutative geometry.

\def\jmp{J.~Math.~Phys.$\,$}
\def\pl{Phys. Lett.$\,$ }
\def\np{Nucl. Phys.$\,$}
\def\ptp{Prog. Theor. Phys.$\,$}
\def\prl{Phys. Rev. Lett.$\,$}
\def\pr{Phys. Rev. D$\,$}
\def\mp{Int. Journ. Mod. Phys.$\,$ }

\end{document}